\documentclass{article}
\usepackage{a4wide}
\usepackage[latin1]{inputenc}
\usepackage{amsmath}
\usepackage{amssymb}
\usepackage{amsfonts}
\usepackage[frenchb]{babel}
\usepackage[T1]{fontenc}
\newcommand{\bol}{\boldsymbol}

\newcommand{\di}{\, \mbox{div}}
\newcommand{\Di}{\, \mbox{\textbf{div}}}

\newcommand{\dev}[1]{\mbox{\textbf{dev}}\left(#1\right)}

\begin{document}
\begin{center}
\framebox{
\begin{minipage}{150mm}
\large{Un lecteur attentif --~que je tiens ici \`a remercier chaleureusement~-- m'a signal\'e une erreur dans la premi\`ere version de cet article (cf. Eq.\,8-2), d\'epos\'ee sur le site HAL le 9 octobre 2008. Il \'etait d'autant plus indispensable que je la corrige que ses cons\'equences sont profondes. Ainsi, dans cette nouvelle version --~o\`u j'ai \'egalement renonc\'e \`a consid\'erer la masse volumique partielle d'une des deux esp\`eces comme variable d'\'etat~--, la premi\`ere loi de Fick a-t-elle un caract\`ere beaucoup plus g\'en\'eral que dans la premi\`ere version.}
\end{minipage}
}
\end{center}
\vspace*{10mm}
\begin{center}
\textbf{\large{Sur la premi\`ere loi de Fick et les \'equations de mouvements quasi-statiques gouvernant la diffusion mol\'eculaire.}}\\
\vspace{4mm}
\normalsize
Thierry D\'ESOYER (thierry.desoyer@ec-marseille.fr), EC-Marseille \& LMA (UPR 7051-CNRS),\\
Technop\^ole de Ch\^ateau-Gombert, 38 rue Joliot Curie, 13451 Marseille Cedex 20, France.
\end{center}
\vspace{10mm}
\textbf{\large{R\'esum\'e}}\\
Le probl\`eme de la diffusion mol\'eculaire dans un m\'elange fluide bi-esp\`ece est ici abord\'e des deux points de vue compl\'ementaires de la M\'ecanique des milieux continus --~d'une mani\`ere quelque peu diff\'erente de celle retenue par Truesdell dans [1]~-- et de la Thermodynamique. Il est ainsi montr\'e que la force li\'ee \`a la 'tra\^in\'ee diffusive', \textit{i.e.}\,au frottement visqueux inter-constituant, est n\'ecessairement li\'ee \`a la vitesse de diffusion relative d'un constituant par rapport à l'autre. La premi\`ere loi de Fick est \'egalement retrouv\'ee en tant que condition suffisante \`a la positivit\'e de la dissipation  associ\'ee aux mouvements diffusifs, laquelle ne peut cependant \^etre \'etablie que si, au pr\'elable, les \'equations r\'egissant les mouvements diffusifs quasi-statiques sont elles-m\^emes \'etablies.\\
\\
\small{\it Mots-cl\'es :} M\'ecanique des milieux continus ; mouvement diffusif ; principe fondamental de la dynamique ; compatibilit\'e thermodynamique ; premi\`ere loi de Fick.\\
\\
\\
\\
\textbf{\large{Abstract}}\\
\normalsize
The problem of the molecular diffusion in a biphasic fluid mixture is studied here from the two complementary points of view of Continuum Mechanics --~in a somewhat different manner from Truesdell in [1]~--\;and that of Thermodynamics. It is established that the force involved in the 'diffusive drag', \textit{i.e.} in the inter-constituent viscous friction, is necessarily linked to the relative diffusion velocity of one constituent with respect to the other. We also end up with Fick's first law, which appears to be a sufficient condition for the dissipated power associated with the diffusive motions to be positive.
\\
\\
\\
\section{Introduction}
\label{sec:intro}
Il est largement admis que la premi\`ere loi de Fick --~~ainsi que la seconde, d'ailleurs, qui n'est cependant rien d'autre que la combinaison de la premi\`ere avec l'\'equation de conservation de la masse de l'esp\`ece consid\'er\'ee~-- d\'ecrit correctement le ph\'enom\`ene de diffusion mol\'eculaire, au moins dans certaines circonstances (\'ecoulements non turbulents de m\'elanges fluides compos\'es de deux esp\`eces \'electriquement et chimiquement neutres, p. ex.). Cette loi, cependant, ne repose initialement sur
\footnote{Les traductions de l'article de Truesdell ici propos\'ees sont de moi.}
 \og... \textit{aucune base} [physique] \textit{sinon une analogie av\'er\'ee avec l'\'ecoulement} [\textit{sic}] \textit{de la chaleur}..."', ainsi que le mentionne Truesdell dans [1].\\
Pour combler cette lacune, Truesdell, toujours dans [1], a propos\'e une th\'eorie m\'ecanique de la diffusion à la fois rigoureuse et pertinente. Celle-ci est essentiellement bas\'ee sur l'id\'ee que \og... \textit{les tra\^in\'ees diffusives sont à l'origine de la force g\'en\'erant les mouvements} [diffusifs]...\fg. Plus pr\'ecis\'ement, l'approche de Truesdell est bas\'ee sur la notion de 'momentum supply'
\footnote{Je n'ai pas trouv\'e de traduction satisfaisante de cette expression.}
, laquelle r\'esulte, dans chacune des esp\`eces du m\'elange, \og... \textit{d'un surcro\^it de forces agissant sur l'esp\`ece par rapport aux forces ext\'erieures} [la gravit\'e, p. ex.] \textit{appliqu\'ees \`a l'esp\`ece et aux forces g\'en\'er\'ees par son contact avec son environnement} [les contraintes].\fg\\
Donnant une expression particuli\`ere \`a ce 'surcro\^it de forces', Truesdell d\'emontre ensuite un certain nombre de r\'esultats importants. Sans les remettre en cause en quoi que ce soit, on peut toutefois souligner que cette expression du 'surcro\^it de forces' qui les sous-tend est postul\'ee \textit{a priori} (en coh\'erence avec l'id\'ee premi\`ere que les 'tra\^in\'ees diffusives' sont \`a l'origine des forces, bien \'evidemment). Il n'est alors pas ill\'egitime de se demander si cette expression est la seule physiquement admissible ou si des alternatives peuvent lui \^etre trouv\'ees : c'est à cette question que, en premier lieu, le pr\'esent article entend donner quelques \'el\'ements de r\'eponse. Il faut toutefois tout de suite souligner que les r\'esultats que l'on donne ici ne sont valables que pour un fluide bi-esp\`ece, alors que ceux \'etablis par Truesdell dans [1] concernent des m\'elanges constitu\'es d'un nombre quelconque d'esp\`eces. Il est tout aussi important de signaler d\`es maintenant que, \`a la diff\'erence de Truesdell, on ne base pas cette \'etude sur les \'equations de mouvement ('absolu') de chacune des esp\`eces mais sur les \'equations de leur mouvement diffusif, lequel n'est que relatif. La notion de 'surcro\^it de forces' se trouve ainsi remplac\'ee par celle 
\footnote{Ces deux notions sont \'equivalentes. De mon point de vue, toutefois, celle de force de frottement inter-esp\`ece est plus facile \`a interpr\'eter physiquement.}
de force de frottement inter-esp\`ece.\\
Cet article est organis\'e comme suit : les notions m\'ecaniques strictement n\'ecessaires \`a l'\'etude sont pr\'esent\'ees dans le le Paragraphe\,(\ref{sec:hyp}), o\`u les importantes \'equations d'\'equilibre (quasi-statique) diffusif sont \'egalement \'etablies. Le Paragraphe\,(\ref{sec:Thermo}) est tout d'abord consacr\'e aux premier et second principes de la Thermodynamique tels qu'ils s'expriment compte tenu des \'equations m\'ecaniques \'etablies dans Para.\,(\ref{sec:hyp}). Des conditions n\'ecessaires et des conditions suffisantes \`a la v\'erification syst\'ematique de l'in\'egalit\'e de Clausius-Duhem y sont \'egalement propos\'es qui, notamment, permettent de pr\'eciser l'expression de la force de frottement inter-esp\`eces et de retrouver la premi\`ere loi de Fick.
\section{Hypoth\`eses g\'en\'erales et \'equations d'\'equilibre quasi-statique}
\label{sec:hyp}
Le syst\`eme mat\'eriel fini (ou global) consid\'er\'e dans cet article est un m\'elange fluide compos\'e de deux esp\`eces, chimiquement et \'electriquement neutres. Le domaine occup\'e par ce syst\`eme \`a l'instant g\'en\'erique $t$ est not\'e $D$ ($D\subset\mathbb{R}^3$). Un quelconque sous-domaine infinit\'esimal de $D$ est indiff\'eremment appel\'e 'particule' ou 'syst\`eme local' par la suite. La description d'Euler est retenue pour tous les champs.\\
La temp\'erature est suppos\'ee \^etre localement la m\^eme dans les deux esp\`eces. Le jeu de variables d'\'etat caract\'erisant l'\'etat thermom\'ecanique de la particule peut alors \^etre restreint \`a :
\begin{equation}
\label{eq:vareta}
T, \rho, \rho_1
\end{equation}
o\`u $T$ est la temp\'erature absolue et $\rho$ (resp. $\rho_1$) la masse volumique du m\'elange (resp. la masse volumique partielle de l'esp\`ece  $1$). La masse volumique partielle de l'esp\`ece $2$ est li\'ee \`a celle de l'esp\`ece $1$ et \`a la masse volumique du m\'elange par :
\begin{equation}
\label{eq:relrho}
\rho\,=\,\sum_{k=1}^2\rho_k
\end{equation}
La premi\`ere hypoth\`ese g\'en\'erale de cette \'etude est relative \`a la puissance des efforts int\'erieurs \`a l'\oe uvre dans le syst\`eme local. Consid\'erant que :\\
i\;--~(cf. \'egalement H2-1)\,l'\'etude du seul mouvement moyen, auquel une puissance des efforts int\'erieurs est associ\'ee, est \'evidemment insuffisante pour rendre compte de la principale caract\'eristique du ph\'enom\`ene de diffusion mol\'eculaire, soit le mouvement diffusif de chacune des deux esp\`eces,\\
ii\;--~(cf. \'egalement H2-2)\,en cons\'equence, ces deux mouvements diffusifs doivent \^etre explicitement pris en compte avec, pour chacun, une puissance des efforts int\'erieurs associ\'ee \textit{a priori} non nulle,\\
iii\;--~(cf. \'egalement H2-3)\,le mouvement diffusif relatif de l'esp\`ece $2$ par rapport \`a l'esp\`ece $1$ (ou \textit{vice-versa}) est \`a l'origine de la force de frottement contribuant elles aussi \`a la puissance des efforts int\'erieurs,\\
cette premi\`ere hypoth\`ese s'\'enonce :
\vspace{2mm}
\begin{center}
\textbf{H1}:\,La \emph{puissance des efforts int\'erieurs} locale (par unit\'e de volume) est la somme de :
\begin{itemize}
	\item\,(H1-1)\,la puissance des efforts int\'erieurs associ\'ee au mouvement moyen du m\'elange : $P^{im}$,
	\item\,(H1-2)\,la puissance des efforts int\'erieurs associ\'ee au mouvement diffusif de chacune des deux esp\`eces :
$$
	P^{id}=\sum_{k=1}^2\,P^{id}_k
$$
	\item\,(H1-3)\,la puissance des efforts int\'erieurs associ\'ee au mouvement diffusif relatif de l'esp\`ece $2$ par rapport \`a l'esp\`ece $1$ (ou \textit{vice-versa}): $P^{ir}$
\end{itemize}
\end{center}
\vspace*{2mm}
D'apr\`es \textbf{H1}, la puissance volumique des efforts int\'erieurs, $P^i$, s'\'ecrit donc :
\begin{equation}
\label{eq:pinttot}
P^i\,=\,P^{im}+\sum_{k=1}^2\,P^{id}_k+P^{ir}
\end{equation}
Dans la seconde hypoth\`ese g\'en\'erale de cette \'etude, chacun des trois termes apparaissant dans \textbf{H1} est classiquement d\'efini comme le produit scalaire d'une variable de type force et d'une variable cin\'ematique. Le principe d'\emph{indiff\'erence mat\'erielle} (cf., p. ex., Truesdell and Noll, [2] ; on parlera ici d'objectivit\'e plut\^ot que d'indiff\'erence mat\'erielle, m\^eme si ces deux auteurs ont finalement renonc\'e \`a utiliser ce terme) fait que :\,i\;--~puisque la vitesse moyenne,\,$\bol{v}^m$, n'est pas objective, il est impossible de la faire intervenir dans l'expression de $P^{im}$. En revanche, la partie sym\'etrique de son gradient eul\'erien, $\bol{\nabla}^s(\bol{v}^m)$, est bien objective, que l'on associera classiquement \`a un tenseur des contraintes moyennes, $\bol{\sigma}^m$, de façon \`a ce que $P^{im}$ soit objectivement d\'efinie ; ii\;--~de la m\^eme façon, pour chacune des deux esp\`eces, on \'ecrira que $P^{id}_k$ est le produit scalaire d'un tenseur des contraintes de diffusion, $\bol{\sigma}^d_k$, et de la partie sym\'etrique du gradient eul\'erien des vitesses de diffusion $\bol{v}^d_k$, $\bol{\nabla}^s(\bol{v}^d_k)$. On supposera cependant, en toute premi\`ere approximation, que ce tenseur des contraintes de diffusion est purement sph\'erique, \textit{i.e.} qu'il se r\'eduit \`a une pression de diffusion, $-p^d_k\bol{G}$, o\`u $\bol{G}$ d\'esigne le tenseur m\'etrique ; iii\;--~quant \`a $P^{ir}$, on la d\'efinira comme le produit scalaire de la vitesse de diffusion relative, $\bol{v}^r=\bol{v}^d_2-\bol{v}^d_1$, qui est bien objective, et d'un vecteur force (par unit\'e de volume) de frottement inter-esp\`eces, $\bol{f}^r$.\\
En r\'esum\'e :
\vspace{2mm}
\begin{center}
\textbf{H2}:\,les variables locales \emph{cin\'ematiques} et \emph{sth\'eniques} intervenant dans le probl\`eme sont :
\begin{itemize}
	\item\,(H2-1)\,la vitesse moyenne du m\'elange ($\bol{v}^m$; non objective), la partie sym\'etrique de son gradient eul\'erien  ($\bol{D}^m\,\hat{=}\,\bol{\nabla}^s(\bol{v}^m)$; objective) et un tenseur des contraintes moyennes ($\bol{\sigma}^m=-p^m\bol{G}+\dev{\bol{\sigma}^m}$, qui devra \^etre objectif), tels que :
$$
-P^{im}\,=\,\bol{\sigma}^m\bol{:}\bol{D}^m\,=\,-p^m\bol{G}\bol{:}\bol{D}^m+\dev{\bol{\sigma}^m}\bol{:}\dev{\bol{D}^m}
$$
	\item\,(H2-2)\,pour chacune des deux esp\`eces, la vitesse de diffusion ($\bol{v}^d_k$; objective), la partie sym\'etrique de son gradient eul\'erien ($\bol{D}^d_k\,\hat{=}\,\bol{\nabla}^s(\bol{v}^d_k)$; objective) et un tenseur des contraintes de diffusion sph\'erique, \textit{i.e.} une pression de diffusion ($\bol{\sigma}^d_k\,=\,-p^d_k\bol{G}$, qui devra \^etre objectif), tels que :
$$
\bol{v}^d_k\,=\,\bol{v}_k\,-\,\bol{v}^m\;\;\;\;;\;\;\;\,-P^{id}_k\,=\,\bol{\sigma}^d_k\bol{:}\bol{D}^d_k\,=\,-p^d_k\di(\bol{v}^d_k)
$$
o\`u $\bol{v}_k$ est la vitesse de l'esp\`ece $k$ (non objective). Il est \`a noter que $\bol{v}^d_k\,=\,\bol{v}_k$ quand $\bol{v}^m=0$ (m\'elange au repos). Il est \'egalement \`a noter que, \`a ce niveau de l'\'etude, aucune relation n'a \`a \^etre impos\'ee, ni entre $\bol{v}^m$ et les vitesses des esp\`eces, $\bol{v}_k$, ni entre les vitesses de diffusion, $\bol{v}^d_k$.
	\item\,(H2-3)\,la vitesse de diffusion relative de l'esp\`ece $2$ par rapport \`a l'esp\`ece $1$ ($\bol{v}^r$; objective) et une force volumique de frottement inter-esp\`eces ($\bol{f}^r$, qui devra \^etre objective), tels que :
$$
\bol{v}^r\,=\,\bol{v}^d_2-\bol{v}^d_1\;\;\;\;;\;\;\;\,-P^{ir}\,=\,\bol{f}^r\bol{.}\bol{v}^r\,=\,\bol{f}^r\bol{.}(\bol{v}^d_2-\bol{v}^d_1)\;\;\;\;\;\,(\textrm{Rmq.:}\;\,\bol{v}^d_2-\bol{v}^d_1\,=\,\bol{v}_2-\bol{v}_1)
$$
\end{itemize}
\end{center}
\vspace*{2mm}
On restreindra d\'esormais l'\'etude aux seuls mouvements, tant moyens que diffusifs, quasi-statiques, soient $\dot{\bol{v}}^m\approx 0$, $\dot{\bol{v}}^d_1\approx 0$, $\dot{\bol{v}}^d_2\approx 0$, la notation $\dot{\bol{a}}$, quelle que soit la grandeur scalaire ou tensorielle $\bol{a}$, d\'esignant la d\'eriv\'ee particulaire de $\bol{a}$ suivant le mouvement moyen du m\'elange, \textit{i.e.} :
\begin{equation}
\label{eq:derpar}
\dot{\bol{a}}\,=\,\frac{\partial\bol{a}}{\partial t}+\bol{\nabla}\bol{a}\bol{.}\bol{v}^m
\end{equation}
La puissance des quantit\'es d'acc\'el\'eration \'etant ainsi n\'egligeable, la troisi\`eme hypoth\`ese g\'en\'erale de l'\'etude, relative \`a la puissance des efforts ext\'erieurs, s'\'enonce comme suit :
\vspace{2mm}
\begin{center}
\textbf{H3}:\,La \emph{puissance des efforts ext\'erieurs} est la somme de ( $\Omega$ est un quelconque sous-domaine du domaine $D$ occup\'e par le fluide) : 
\begin{itemize}
	\item\,(H3-1)\,la puissance des efforts ext\'erieurs associ\'ee au mouvement moyen, telle que ($\bol{g}$ est l'acc\'el\'eration de la pesanteur et $\bol{F}^m$ une force surfacique) :
$$
\mathbb{P}^{em}=\int_{\Omega}\rho\,\bol{g}\bol{.}\bol{v}^mdV+\int_{\partial\Omega}\bol{F}^m\bol{.}\bol{v}^mdS
$$
	\item\,(H3-2)\,pour chacune des deux esp\`eces, la puissance des efforts ext\'erieurs associ\'ee au mouvement diffusif, telle que ($\bol{F}^{dk}$ est une force surfacique) :
$$
\mathbb{P}^{ed}_k\,=\,\int_{\partial\Omega}\bol{F}^{dk}\bol{.}\bol{v}^d_k\,dS\;\;\;\Rightarrow\;\;\,\mathbb{P}^{ed}=\,\sum_{k=1}^2\mathbb{P}^{ed}_k
$$
	\item\,(H3-3)\,la puissance des efforts ext\'erieurs associ\'ee au mouvement diffusif relatif, telle que :
$$
\mathbb{P}^{er}\,=\,0
$$
\end{itemize}
\end{center}
\vspace{2mm}
La puissance totale des efforts ext\'erieurs, $\mathbb{P}^e$, d'apr\`es \textbf{H3}, s'\'ecrit donc :
\begin{equation}
\label{eq:pexttot}
\mathbb{P}^e\,=\,\mathbb{P}^{em}+\sum_{k=1}^2\,\mathbb{P}^{ed}_k
\end{equation}
Puisque le principe fondamental de la dynamique (restreint ici, on le rappelle, \`a des mouvements quasi-statiques) :
\begin{equation}
\label{eq:pfd}
\mathbb{P}^i+\mathbb{P}^e=0\;\;\;\textrm{avec}\;\;\,\mathbb{P}^i=\int_{\Omega}P^i\,dV
\end{equation}
doit \^etre satisfait quels que soient $\bol{v}^m$, $\bol{v}^d_1$, $\bol{v}^d_2$ et $\Omega\subset D$, les classiques \'equations d'\'equilibre quasi-statique et conditions aux limites sont retrouv\'ees pour le mouvement quasi-statique moyen, soient :
\begin{equation}
\label{eq:eqmoy}
\Di(\bol{\sigma}^m)+\rho\bol{g}\,=\,0\;\,\textrm{dans}\;D\;\;\,;\;\;\bol{\sigma}^m\bol{.}\bol{N}\,=\,\bol{F}^m\;\,\textrm{sur}\;\partial D
\end{equation}
tandis que, pour le mouvement diffusif quasi-statique de chacune des deux esp\`eces, ces \'equations s'\'ecrivent :
\begin{equation}
\label{eq:eqdif}
\bol{\nabla}p^d_k+\theta_k\bol{f}^r\,=\,0\;\,\textrm{dans}\;D\;\;\,;\;\;-p^d_k\bol{N}\,=\,\bol{F}^{dk}\;\,\textrm{sur}\;\partial D
\end{equation}
o\`u $\theta^1=-1$ et $\theta^2=1$. En cons\'equence :
\begin{equation}
\label{eq:eqdif2}
\sum_{k=1}^2\,\left(\bol{\nabla}p^d_k+\theta_k\bol{f}^r\right)\,=\,\sum_{k=1}^2\,\bol{\nabla}p^d_k\,=\,0\;\,\textrm{dans}\;D
\end{equation}
Les conditions aux limites Eq.\,\eqref{eq:eqdif}-2 sont sans int\'er\^et pour la suite de l'\'etude. On ne les commentera donc pas dans cet article.\\
\section{Compatibilit\'e thermodynamique : hypoth\`ese, conditions n\'ecessaires et conditions suffisantes}
\label{sec:Thermo}
Le potentiel d'\'etat d'\'energie interne massique (resp. d'\'energie libre massique) du syst\`eme local est not\'e $e(T, \rho,\rho_1)$ (resp. $\psi(T, \rho,\rho_1)$), avec $e=\psi+sT$, o\`u $s$ est l'entropie massique. Le premier principe de la thermodynamique s'\'ecrit simplement (cf. p. ex. Garrigues, [3]):
\begin{equation}
\label{eq:pt1-1}
\rho\,\dot{e}\,=\,-\di(\bol{q})-P^i\;\;\;\Leftrightarrow\;\;\,\rho\,\dot{\psi}+\rho\,T\dot{s}+\rho\,s\dot{T}\,=\,-\di(\bol{q})-P^i
\end{equation}
o\`u $\bol{q}$ est le vecteur flux de chaleur et o\`u la puissance des efforts int\'erieurs $P^i$, cf. Eq.\,\eqref{eq:pinttot}, est la somme des puissances associées au mouvement moyen du m\'elange, au mouvement diffusif de chacune des esp\`eces et au mouvement diffusif relatif d'une esp\`ece par rapport \`a l'autre, soit, selon \textbf{H1} :
\begin{equation}
\label{eq:pint1}
-P^i\,=\,-p^m\,\bol{G}\bol{:}\bol{D}^m\,+\,\dev{\bol{\sigma}^m}\bol{:}\dev{\bol{D}^m}\,-\,p^d_1\,\bol{G}\bol{:}\bol{D}^d_1\,-\,p^d_2\,\bol{G}\bol{:}\bol{D}^d_2\,+\,\bol{f}^r\bol{.}(\bol{v}^d_2-\bol{v}^d_1)
\end{equation}
Par ailleurs, l'\'equation de conservation de la masse du m\'elange s'\'ecrit :
\begin{equation}
\label{eq:conmas-mel}
\dot{\rho}\,=\,-\rho\,\bol{G}\bol{:}\bol{D}^m
\end{equation}
alors que celle de l'esp\`ece $1$ s'\'ecrit, en accord avec la relation liant la vitesse de l'esp\`ece $1$, sa vitesse de diffusion et la vitesse du m\'elange (cf. H2-2) :
\begin{equation}
\label{eq:conmas-esp1}
\frac{\;\partial\,\rho_1}{\partial\, t}+\bol{\nabla}\rho_1\bol{.}\bol{v}_1\,=\,-\rho_1\di(\bol{v}_1)\,=\,-\rho_1\di(\bol{v}^m+\bol{v}^d_1)\,=\,-\rho_1\bol{G}\bol{:}(\bol{D}^m+\bol{D}^d_1)
\end{equation}
ou encore, puisque, selon Eq.\,\eqref{eq:derpar}, $\dot{\rho}^1\,=\,\partial\rho_1/\partial t+\bol{\nabla}\rho_1\bol{.}\bol{v}^m$ :
\begin{equation}
\label{eq:conmas-esp2}
\dot{\rho}^1\,=\,-\rho_1\bol{G}\bol{:}(\bol{D}^m+\bol{D}^d_1)-\bol{\nabla}\rho_1\bol{.}\bol{v}^d_1
\end{equation}
Il est ici \`a noter que les \'equations Eq.\,\eqref{eq:conmas-esp2} sont compatibles avec celle issue de Eq.\,\eqref{eq:relrho}, $\dot{\rho}=\dot{\rho}^1+\dot{\rho}^2$, si et seulement si la relation usuelle entre les deux vitesses de diffusion est v\'erifi\'ee :
\begin{equation}
\label{eq:vitdif}
\rho_1\bol{v}^d_1+\rho_2\bol{v}^d_2=0
\end{equation}
Si l'on retient alors l'hypoth\`ese d'un \og m\'elange binaire \`a masse volumique uniforme \fg, dont Truesdell rappelle dans [1] qu'elle est n\'ecessaire \`a la validit\'e de la loi de Fick, soit :
\vspace{2mm}
\begin{center}
\textbf{H4} :\,$\bol{\nabla}\rho\,=\,0\;\;\,\Leftrightarrow\;\;\,\bol{\nabla}\rho_2\,=\,-\,\bol{\nabla}\rho_1$
\end{center}
on d\'eduit imm\'ediatement de Eq.\,\eqref{eq:vitdif} que :
\begin{equation}
\label{eq:gradrhok}
\bol{\nabla}\bol{v}^d_2=\,-\,\frac{1}{\rho_2}\left(\frac{\rho}{\rho_2}\,\bol{\nabla}\rho_1\otimes\bol{v}^d_1+\rho_1\bol{\nabla}\bol{v}^d_1\right)\;\;\Rightarrow\;\;\bol{G}\bol{:}\bol{D}^d_2=\,-\,\frac{1}{\rho_2}\left(\frac{\rho}{\rho_2}\,\bol{\nabla}\rho_1\bol{.}\bol{v}^d_1+\rho_1\bol{G}\bol{:}\bol{D}^d_1\right)
\end{equation}
Une nouvelle expression de la puissance volumique des efforts int\'erieurs, cf. Eq.\,\eqref{eq:pint1}, est ainsi obtenue \`a partir de Eq.\,\eqref{eq:vitdif} et Eq.\,\eqref{eq:gradrhok}, soit :
\begin{equation}
\label{eq:pint2}
-\,P^i\,=\,-p^m\,\bol{G}\bol{:}\bol{D}^m\,+\,\dev{\bol{\sigma}^m}\bol{:}\dev{\bol{D}^m}\,+\,\left(-p^d_1+\frac{\rho_1}{\rho_2}p^d_2\right)\bol{G}\bol{:}\bol{D}^d_1\,+\,\left(\frac{\rho}{\rho_2^2}\,p^d_2\,\bol{\nabla}\rho_1\,-\,\frac{\rho}{\rho_2}\,\bol{f}^r\right)\bol{.}\bol{v}^d_1
\end{equation}
\`A partir de \textbf{H1}, \textbf{H2} et Eq.\,\eqref{eq:conmas-esp2}, Eq.\,\eqref{eq:pt1-1} peut alors \^etre r\'ecrite :
$$
\rho\,T\dot{s}+\di(\bol{q})\,=\,-\rho\left(s+\frac{\partial\psi}{\partial T}\right)\dot{T}+\left(-p^m+\rho^2\,\frac{\partial\psi}{\partial\rho}+\rho\rho_1\,\frac{\partial\psi}{\partial\rho_1}\right)\bol{G}\bol{:}\bol{D}^m+\dev{\bol{\sigma}^m}\bol{:}\dev{\bol{D}^m}
$$
\begin{equation}
\label{eq:pt1-2}
+\,\left(-p^d_1+\frac{\rho_1}{\rho_2}\,p^d_2+\rho\rho_1\,\frac{\partial\psi}{\partial\rho_1}\right)\bol{G}\bol{:}\bol{D}^d_1\,+\,\left(\left(\rho\,\frac{\partial\psi}{\partial\rho_1}+\frac{\rho}{\rho_2^2}\,p^d_2\right)\bol{\nabla}\rho_1-\,\frac{\rho}{\rho_2}\,\bol{f}^r\right)\bol{.}\bol{v}^d_1
\end{equation}
Le second principe de la Thermodynamique, quant \`a lui, doit \^etre v\'erifi\'e (cf. Garrigues, [3]): i\;--~quel que soit l'\'etat de la particule, $(T,\rho,\rho_1)$;\;ii\;--~quelle que soit l'\'evolution subie par la particule, $(\,\dot{T},\bol{D}^m,\bol{D}^d_1,\bol{v}^d_1)$;\; iii\;--~quel que soit le gradient de temp\'erature agissant sur la particule, $\bol{\nabla}T$, \textit{i.e.}:
\begin{equation}
\label{eq:pt2-1}
\rho\,T\dot{s}+\di(\bol{q})-\frac{1}{T}\bol{q}\bol{.}\bol{\nabla}T\geq 0\;\;\;\;\;\forall(T, \rho_1,\rho_2)\,,\;\forall(\dot{T},\bol{D}^m,\bol{D}^d_1,\bol{v}^d_1)\,,\;\forall\,\bol{\nabla}T
\end{equation}
En combinant Eq.\,\eqref{eq:pt1-2} et Eq.\,\eqref{eq:pt2-1}, on obtient alors l'in\'egalit\'e de Clausius-Duhem (qui stipule que la dissipation $\phi$, \textit{i.e.} la puissance volumique dissip\'ee, est n\'ecessairement non n\'egative), soit :
$$
\phi\,=\,-\frac{1}{T}\bol{q}\bol{.}\bol{\nabla}T\,-\rho\left(s+\frac{\partial\psi}{\partial T}\right)\dot{T}+\left(-p^m+\rho^2\,\frac{\partial\psi}{\partial\rho}+\rho\rho_1\,\frac{\partial\psi}{\partial\rho_1}\right)\bol{G}\bol{:}\bol{D}^m+\dev{\bol{\sigma}^m}\bol{:}\dev{\bol{D}^m}
$$
$$
+\,\left(-p^d_1+\frac{\rho_1}{\rho_2}\,p^d_2+\rho\rho_1\,\frac{\partial\psi}{\partial\rho_1}\right)\bol{G}\bol{:}\bol{D}^d_1\,+\,\left(\left(\rho\,\frac{\partial\psi}{\partial\rho_1}+\frac{\rho}{\rho_2^2}\,p^d_2\right)\bol{\nabla}\rho_1-\,\frac{\rho}{\rho_2}\,\bol{f}^r\right)\bol{.}\bol{v}^d_1\geq 0
$$
\begin{equation}
\label{eq:pt2-2}
\forall(T,\rho,\rho_1)\,,\;\forall(\dot{T},\bol{D}^m,\bol{D}^d_1,\bol{v}^d_1)\,,\;\forall\,\bol{\nabla}T
\end{equation}
Si l'on suppose alors que:
\vspace{2mm}
\begin{center}
\textbf{H5} :\,aucune des "inconnues" (soient $\bol{q}$, $p^m$, $\dev{\bol{\sigma}^m}$, $p^d_k$, $\bol{\nabla}\rho_1$ et $\bol{f}^r$)\\apparaissant dans Eq.\,\eqref{eq:pt2-2} ne d\'epend de $\dot{T}$
\end{center}
le fait que l'entropie massique $s$ soit une fonction d'\'etat permet alors d'\'etablir une premi\`ere condition \emph{n\'ecessaire} (et trivialement suffisante) \`a la v\'erification syst\'ematique de cette in\'egalit\'e, soit : 
\begin{equation}
\label{eq:entropie}
s\,=\,-\,\frac{\partial\psi}{\partial T}
\end{equation}
L'in\'egalit\'e Eq.\,\eqref{eq:pt2-2} associ\'ee \`a la condition Eq.\,\eqref{eq:entropie} doit \'egalement \^etre v\'erifi\'ee quand $\bol{D}^m=0$ et $\bol{\nabla}T=0$. D'o\`u, n\'ecessairement :
$$
\phi^d_1\,=\,\left(-p^d_1+\frac{\rho_1}{\rho_2}\,p^d_2+\rho\rho_1\,\frac{\partial\psi}{\partial\rho_1}\right)\bol{G}\bol{:}\bol{D}^d_1\,+\,\left(\left(\rho\,\frac{\partial\psi}{\partial\rho_1}+\frac{\rho}{\rho_2^2}\,p^d_2\right)\bol{\nabla}\rho_1-\,\frac{\rho}{\rho_2}\,\bol{f}^r\right)\bol{.}\bol{v}^d_1\geq 0
$$
\begin{equation}
\label{eq:pt2-3}
\forall(T,\rho,\rho_1)\,,\;\forall(\bol{D}^d_1,\bol{v}^d_1)
\end{equation}
Dans Eq.\,\eqref{eq:pt2-3}, seules les variables caract\'eristiques de l'\'evolution de l'esp\`ece $1$, $\bol{D}^d_1$ et $\bol{v}^d_1$ interviennent. De ce fait, $\phi^d_1$ est donc la dissipation volumique associ\'ee au mouvement diffusif de l'esp\`ece $1$.\\
Si l'on admet alors que:
\vspace{2mm}
\begin{center}
\textbf{H6} :\,les pressions de diffusion, $p^d_k$, ne d\'ependent que des variables d'\'etat
\end{center}
une premi\`ere condition \emph{n\'ecessaire} (et trivialement suffisante) \`a la v\'erification syst\'ematique de l'in\'egalit\'e Eq.\,\eqref{eq:pt2-3}, est :
\begin{equation}
\label{eq:pt2-4}
p^d_1\,=\,\frac{\rho_1}{\rho_2}\,p^d_2+\rho\rho_1\,\frac{\partial\psi}{\partial\rho_1}
\end{equation}
De Eq.\,\eqref{eq:pt2-4}, en accord avec \textbf{H4} et sachant que, d'apr\`es Eq.\,\eqref{eq:eqdif}, $\bol{\nabla}p^d_1=-\bol{\nabla}p^d_2=\bol{f}^r$, on d\'eduit alors que :
\begin{equation}
\label{eq:pt2-5}
\frac{\rho}{\rho_2}\,\bol{f}^r\,=\,\frac{\rho}{\rho_2^2}\,p^d_2\,\bol{\nabla}\rho_1\,+\,\bol{\nabla}\left(\rho\,\rho_1\,\frac{\partial\psi}{\partial\rho_1}\right)
\end{equation}
ou encore, toujours en accord avec \textbf{H4} et sachant qu'on a suppos\'e $\bol{\nabla}T=0$ :
\begin{equation}
\label{eq:pt2-6}
\frac{\rho}{\rho_2}\,\bol{f}^r\,=\,\left(\frac{\rho}{\rho_2^2}\,p^d_2\,+\,\rho\,\frac{\partial\psi}{\partial\rho_1}\,+\,\rho\rho_1\,\frac{\partial^2\psi}{\partial\rho_1^2}\right)\bol{\nabla}\rho_1
\end{equation}
Il appara\^it alors que, d'apr\`es Eq.\,\eqref{eq:pt2-4}et Eq.\,\eqref{eq:pt2-6}, la dissipation volumique associ\'ee au mouvement diffusif de l'esp\`ece $1$, cf. Eq.\,\eqref{eq:pt2-3}, peut s'exprimer en fonction des variables d'\'etat, de la vitesse de diffusion de l'esp\`eces $1$ et du seul gradient de masse volumique partielle de cette esp\`ece, soit :
\begin{equation}
\label{eq:pt2-7}
\phi^d_1\,=\,-\,\rho\rho_1\,\frac{\partial^2\psi}{\partial\rho_1^2}\bol{\nabla}\rho_1\bol{.}\bol{v}^d_1\geq 0\;\;\;\;\,\forall(T,\rho,\rho_1)\,,\;\forall(\bol{D}^d_1,\bol{v}^d_1)
\end{equation}
Une condition \emph{n\'ecessaire et suffisante} \`a la v\'erification syst\'ematique de Eq.\,\eqref{eq:pt2-7} est que $\bol{\nabla}\rho_1$ soit une fonction, $\bol{h}_1$, d\'ependant explicitement de $\bol{v}^d_1$, soit, en toute g\'en\'eralit\'e :
\begin{equation}
\label{eq:cnsgrad}
\bol{\nabla}\rho_1\,=\,sgn\left(-\,\rho\rho_1\,\frac{\partial^2\psi}{\partial\rho_1^2}\right)\,\bol{h}_1(T,\rho,\rho_1,\bol{D}^d_1,\bol{v}^d_1)\;\;\;\;\textrm{avec}\;\;\;\;\bol{h}_1(T,\rho,\rho_1,\bol{D}^d_1,\bol{v}^d_1)\bol{.}\bol{v}^d_1\geq 0\;\;\;\forall(T,\rho,\rho_1)\,,\;\forall(\bol{D}^d_1,\bol{v}^d_1)
\end{equation}
o\`u $sgn(x)=+1$ si $x>0$ et $sgn(x)=-1$ si $x<0$. L'hypoth\`ese formul\'ee par Truesdell --~les forces dues \`a la tra\^in\'ee diffusive (dont la r\'esultante est repr\'esent\'ee par la force volumique de frottement inter-esp\`eces $\bol{f}^r$ dans cette \'etude) sont explicitement fonction des vitesses de diffusion, cf. [1]~-- se trouve ainsi justifi\'ee par Eq.\,\eqref{eq:pt2-6} et Eq.\,\eqref{eq:cnsgrad}. Selon ces deux m\^emes \'equations, $\bol{f}^r$ d\'epend \'egalement des variables d'\'etat.\\
Quant \`a la fonction $\bol{h}_1$, la plus simple expression que l'on puisse lui donner --~qui n'est donc qu'une condition \emph{suffisante}, parmi d'autres
\footnote{\`A titre d'exemple de condition suffisante alternative \`a Eq.\,\eqref{eq:pt2-8}, \textit{i.e.} \`a la v\'erification syst\'ematique de Eq.\,\eqref{eq:cnsgrad}, on pourrait proposer : $\bol{h}_1\,=\,\,\bol{M}_1\bol{.}(\rho_1\,\bol{v}^d_1)$ o\`u $\bol{M}_1$ serait un tenseur sym\'etrique (de valeurs propres r\'eelles, donc) d\'efini positif, pouvant d\'ependre des variables d'\'etat, de $\bol{v}^d_1$ et de $\bol{D}^d_1$.}
, \`a la v\'erification de l'in\'egalit\'e Eq.\,\eqref{eq:cnsgrad}~-- est :
\begin{equation}
\label{eq:pt2-8}
\bol{h}_1\,=\,\frac{1}{\mu}\,\rho_1\,\bol{v}^d_1\;\;\;\;\,\textrm{avec}\;\;\;\;\,\mu>0
\end{equation}
soit encore, d'apr\`es Eq.\,\eqref{eq:cnsgrad} :
\begin{equation}
\label{eq:pt2-9}
\bol{\nabla}\rho_1\,=\,\frac{1}{\mu}sgn\left(-\,\rho\rho_1\,\frac{\partial^2\psi}{\partial\rho_1^2}\right)\,\rho_1\,\bol{v}^d_1\;\;\;\;\,\textrm{avec}\;\;\;\;\,\mu>0
\end{equation}
Dans Eq.\,\eqref{eq:pt2-8} et Eq.\,\eqref{eq:pt2-9}, $\mu$ (en $m^2.\,s^{-1}$) s'interpr\`ete comme un param\`etre caract\'eristique du frottement visqueux inter-esp\`ece et pouvant \'eventuellement d\'ependre des variables d'\'etat et des variables d'\'evolution : ainsi qu'on l'a d\'ej\`a sp\'ecifi\'e dans Eq.\,\eqref{eq:pt2-8} et Eq.\,\eqref{eq:pt2-9}, la seule contrainte que lui impose la thermodynamique est $\mu>0$. On peut aussi noter que Eq.\,\eqref{eq:pt2-9} est bien coh\'erente avec Eq.\,\eqref{eq:vitdif} et \textbf{H4}.\\
Il appara\^it ainsi finalement que la premi\`ere loi de Fick, soit : 
\begin{equation}
\label{eq:fick}
\bol{\nabla}\rho_1\,=\,\frac{1}{\mu}\,\rho_1\,\bol{v}^d_1\;\;\;\;\,\textrm{avec}\;\;\;\;\,\mu>0
\end{equation}
n'est compatible avec Eq.\,\eqref{eq:pt2-9} que si et seulement si $\partial^2\psi/\partial\rho_1^2<0$. Outre cette restriction, la premi\`ere loi de Fick repose sur plusieurs hypoth\`eses qu'il n'est pas inutile de rappeler ici : $\bol{\nabla}\rho=0$ (cf. \textbf{H4}) ; $\bol{q}$, $p^m$, $\dev{\bol{\sigma}^m}$, $p^d_k$, $\bol{\nabla}\rho_1$ ni $\bol{f}^r$ ne d\'ependent de $\dot{T}$ (cf. \textbf{H5}) ; les pressions de diffusion, $p^d_k$, ne d\'ependent que des variables d'\'etat (cf. \textbf{H6}) ; $\bol{D}^m=0$ et $\bol{\nabla}T=0$ (cf. Rmq.\,1 ci-dessous). Il est \'egalement important de rappeler que cette loi n'est qu'une condition suffisante \`a la non-n\'egativit\'e de la dissipation volumique associ\'ee au mouvement diffusif de l'esp\`ece $1$.\\
\\
En ce qui concerne la partie restante de la dissipation, $\phi^m=\phi-\phi^d_1$, sa non-n\'egativit\'e est assur\'ee \`a diverses conditions \emph{suffisantes}, dont les suivantes (classiques en ce qui concerne $\dev{\bol{\sigma}^m}$) :
\begin{equation}
\label{eq:cspt2}
p^m\,=\,\rho^2\frac{\partial\psi}{\partial\rho}\,+\,\rho\,\rho_1\frac{\partial\psi}{\partial\rho_1}\,\;\;\;\,;\;\;\,\dev{\bol{\sigma}^m}\,=\,\nu\,\dev{\bol{D}^m}
\end{equation}
o\`u $\nu>0$ est la viscosit\'e du m\'elange et $K>0$, sa conductivit\'e thermique. On peut ici remarquer que, d'apr\`es Eq.\,\eqref{eq:pt2-4} et Eq.\,\eqref{eq:cspt2}-2, les pressions moyenne et de diffusion sont li\'ees par :
\begin{equation}
\label{eq:pressions}
p^m\,=\,\rho^2\frac{\partial\psi}{\partial\rho}\,+\,p^d_1\,-\,\frac{\rho_1}{\rho_2}\,p^d_2
\end{equation}
Pour clore ce paragraphe, on soulignera que, de par leurs expressions Eq.\,\eqref{eq:pt2-4}, Eq.\,\eqref{eq:pt2-9} et Eq.\,\eqref{eq:cspt2} en fonction de variables qui sont toutes objectives, $p^d_k$, $\bol{f}^r$, $p^m$ et $\dev{\bol{D}^m}$ sont bien des grandeurs objectives.
\begin{flushleft}
-----
\end{flushleft}
\textsl{--\;Rmq.\,1 :}\;Si l'on renonce \`a l'hypoth\`ese $\bol{\nabla}T=0$ tout en conservant les autres hypoth\`eses, soient $\textbf{H4}$, $\textbf{H5}$, $\textbf{H6}$ et $\bol{D}^m=0$, Eq.\,\eqref{eq:pt2-6} devient :
\begin{equation}
\label{eq:rmq1}
\frac{\rho}{\rho_2}\,\bol{f}^r\,=\,\left(\frac{\rho}{\rho_2^2}\,p^d_2\,+\,\rho\,\frac{\partial\psi}{\partial\rho_1}\,+\,\rho\rho_1\,\frac{\partial^2\psi}{\partial\rho_1^2}\right)\bol{\nabla}\rho_1\,+\,\rho\rho_1\,\frac{\partial^2\psi}{\partial\rho_1\partial T}\,\bol{\nabla}T
\end{equation}
L'expression de la puissance volumique dissip\'ee totale est alors :
\begin{equation}
\label{eq:rmq2}
\phi^d\,=\,-\,\frac{1}{T}\,\bol{q}\bol{.}\bol{\nabla}T\,-\,\rho\rho_1\,\frac{\partial^2\psi}{\partial\rho_1^2}\bol{\nabla}\rho_1\bol{.}\bol{v}^d_1\,-\,\rho\rho_1\,\frac{\partial^2\psi}{\partial\rho_1\partial T}\,\bol{\nabla}T\bol{.}\bol{v}^d_1\geq 0\;\;\;\;\,\forall(T,\rho,\rho_1)\,,\;\forall(\bol{D}^d_1,\bol{v}^d_1)\,,\;\forall\,\bol{\nabla}T
\end{equation}
soit encore, quel que soit $\eta\in[0,1]$ :
$$
\phi^d=-\left(\frac{1}{T}\,\bol{q}+\eta\rho\rho_1\,\frac{\partial^2\psi}{\partial\rho_1\partial T}\,\bol{v}^d_1\right)\bol{.}\bol{\nabla}T\,-\,\rho\rho_1\left(\frac{\partial^2\psi}{\partial\rho_1^2}\bol{\nabla}\rho_1+(1-\eta)\frac{\partial^2\psi}{\partial\rho_1\partial T}\,\bol{\nabla}T\right)\bol{.}\bol{v}^d_1\geq 0
$$
\begin{equation}
\label{eq:rmq3}
\forall(T,\rho,\rho_1)\,,\;\forall(\bol{D}^d_1,\bol{v}^d_1)\,,\;\forall\,\bol{\nabla}T
\end{equation}
Une condition suffisante \`a la v\'erification syst\'ematique de Eq.\,\eqref{eq:rmq3} est (avec $K>0$ et $\mu>0$) :
\begin{equation}
\label{eq:rmq4}
\frac{1}{T}\,\bol{q}+\eta\rho\rho_1\,\frac{\partial^2\psi}{\partial\rho_1\partial T}\,\bol{v}^d_1\,=\,-\frac{K}{T}\bol{\nabla}T\;\;;\;\,sgn\left(-\frac{\partial^2\psi}{\partial\rho_1^2}\right)\bol{\nabla}\rho_1-(1-\eta)\left|-\frac{\partial^2\psi}{\partial\rho_1^2}\right|^{-1}\frac{\partial^2\psi}{\partial\rho_1\partial T}\,\bol{\nabla}T\,=\,\frac{1}{\mu}\rho_1\bol{v}^d_1
\end{equation}
soit encore, en supposant que, comme dans Eq.\,\eqref{eq:fick}, $\partial^2\psi/\partial\rho_1^2<0$ :
$$
\rho_1\bol{v}^d_1\,=\,\mu\bol{\nabla}\rho_1-\left((1-\eta)\mu\left|-\frac{\partial^2\psi}{\partial\rho_1^2}\right|^{-1}\frac{\partial^2\psi}{\partial\rho_1\partial T}\right)\bol{\nabla}T\;\;\;\;;
$$
\begin{equation}
\label{eq:rmq5}
\bol{q}\,=\,-\eta\mu\rho T\,\frac{\partial^2\psi}{\partial\rho_1\partial T}\bol{\nabla}\rho_1\,+\,\left(-K+\eta(1-\eta)\mu\rho T\left|-\frac{\partial^2\psi}{\partial\rho_1^2}\right|^{-1}\left(\frac{\partial^2\psi}{\partial\rho_1\partial T}\right)^2\right)\bol{\nabla}T
\end{equation}
Si $\eta\neq 0$ et $\eta\neq 1$, les \'egalit\'es Eq.\,\eqref{eq:rmq5} traduisent un couplage "thermo-diffusif", lequel peut facilement s'interpr\'eter dans les deux cas limites suivants : i\;--~m\^eme si $\bol{\nabla}T=0$, le flux de chaleur $\bol{q}$ est non nul d\`es que $\bol{\nabla}\rho_1\neq 0$ (effet dit "Dufour"); ii\;--~m\^eme si $\bol{\nabla}\rho_1=0$, la vitesse de diffusion de l'esp\`ece $1$, $\bol{v}^d_1$, est non nulle d\`es que $\bol{\nabla}T\neq 0$ (effet dit "Soret"). La pr\'esentation g\'en\'eralement faite de ces deux effets correspond au cas particulier de Eq.\,\eqref{eq:rmq5} tel que :
\begin{equation}
\label{eq:rmq6}
\frac{1-\eta}{\rho T}\,=\,\eta\left|-\,\frac{\partial^2\psi}{\partial\rho_1^2}\right|
\end{equation}
o\`u les deux termes de l'\'egalit\'e sont strictement positifs puisque qu'il a \'et\'e pr\'ec\'edemment suppos\'e que $\eta\in ]0,1[$ et $\partial^2\psi/\partial\rho_1^2<0$.\\
La premi\`ere loi de Fick, cf. Eq.\,\eqref{eq:fick}, correspond \`a Eq.\,\eqref{eq:rmq5}-1 quand $\bol{\nabla}T=0$. Ainsi qu'on vient de le souligner, elle est associ\'ee \`a un flux de chaleur non nulle. Ce dernier r\'esultat compl\`ete ceux \'etablis dans le Paragraphe \ref{sec:Thermo}.
\begin{flushright}
-----
\end{flushright}
\section{Conclusion}
\label{sec:concl}
Les deux principaux r\'esultats de cette \'etude, valables si et seulement si les mouvements moyen et diffusifs sont quasi-statiques (et, plus anecdotiquement, si les pressions de diffusion sont non visqueuses) et si et seulement si les champs de masse volumique, de temp\'erature et de partie sym\'etrique du gradient eul\'erien des vitesses du m\'elange sont uniformes, sont que : i\;--~l'hypoth\`ese \textit{a priori} de Truesdell, selon laquelle la force associ\'ee \`a la tra\^in\'ee diffusive est explicitement li\'ee \`a la vitesse de diffusion relative, est en fait une condition n\'ecessaire \`a la non-n\'egativit\'e de la dissipation associ\'ee aux mouvements diffusifs ; ii\;--~la premi\`ere loi de Fick, en revanche, n'est qu'une condition suffisante \`a la non-n\'egativit\'e de la dissipation associ\'ee aux mouvements diffusifs, combin\'ee \`a une condition peu contraignante (stricte positivit\'e) sur la d\'eriv\'ee seconde du potentiel d'\'etat d'\'energie libre massique par rapport \`a la masse volumique partielle de l'esp\`ece consid\'er\'ee. Il est \'egalement important de souligner que ces deux conditions n'ont pu \^etre effectivement explicit\'ees dans cette \'etude qu'en tenant compte des \'equations r\'egissant les mouvements diffusifs, lesquelles ont donc d\^u \^etre pr\'elablement \'etablies.\\
\`A titre de perspectives, on peut mentionner que certaines hypoth\`eses formul\'ees dans cette \'etude pourraient \^etre facilement rel\^ach\'ees de façon \`a obtenir une premi\`ere loi de Fick 'g\'en\'eralis\'ee', soient : i\;--~(en lien avec H2-2)\,une partie d\'eviatoire pourrait \^etre ajout\'ee \`a chacune des deux pressions de diffusion ; ii\;--~(en lien avec Eq.\,\eqref{eq:pt2-4}-1) pour chacune des esp\`eces, une partie visqueuse pourrait \^etre ajout\'ee \`a  l'\'etat de pression de diffusion, qui d\'ependrait de la trace de la partie sym\'etrique du gradient eul\'erien de la vitesse de diffusion.\\
Il est finalement \`a noter que l'extension de la d\'emarche suivie dans cette \'etude reste \`a faire : i\;--~pour des m\'elanges constitu\'es de $N>2$ esp\`eces ; ii\;--~pour des mouvements non-quasi-statiques, \textit{i.e.} pour des \'ecoulements de m\'elange et des mouvements diffusifs non quasistatiques et donc potentiellement turbulents.
\newpage
\textbf{\large{R\'ef\'erences bibliographiques}}\\
$[1]$\;C. Truesdell, Mechanical basis of diffusion, J. Chem. Physics (U.S.), 37 (1962) 2336-2344.\\
$[2]$\;C. Truesdell and W. Noll, The non-linear field theories of Mechanics, Vol.III/3, Berlin, Heidelberg, New York, Springer, 1965.\\
$[3]$\;J. Garrigues, Fondements de la m\'ecanique des milieux continus, Herm\`es Science Publications, Paris, 2007.
\end{document}